\documentclass[a4,12pt]{article}  	
	
    \usepackage{abstract} 
    
    \usepackage{titlesec} 

    \usepackage{authblk} 

	\usepackage{datetime} 
    \newdateformat{usvardate}{
  	\monthname[\THEMONTH] \ordinal{DAY}, \THEYEAR}

  	\usepackage[bottom]{footmisc} 
    
    \usepackage{fancyhdr}
    \fancyheadoffset{0cm}
    \usepackage[firstpage]{draftwatermark} 
    \SetWatermarkLightness{0.85}

\newcommand\wordcount{%
  \immediate\write18{texcount -utf8 -merge -sum -incbib -dir -sub=none -brief \jobname.tex | cut -d : -f 1 > 'count.txt'}%
  \input{count.txt}\ignorespaces words%
}

    \usepackage{amssymb}
    \usepackage{amsthm}
    \usepackage{newtxmath}

    \usepackage{microtype} 
    \usepackage[utf8]{inputenc}
    \usepackage{newtxtext} 
  
	\usepackage{setspace} 
    \usepackage{lineno,xcolor} 	

	\usepackage[top=1.5in, bottom=1.5in, left=1in, right=1in]{geometry}

	\usepackage[colorinlistoftodos]{todonotes} 
	\usepackage{soul} 

    \usepackage{graphicx} 
    \usepackage{float} 
    
    \usepackage{printlen}

    \usepackage{longtable} 
    \usepackage{tabularx} 
    \newcolumntype{L}[1]{>{\raggedright\arraybackslash}p{#1}}
    \newcolumntype{C}[1]{>{\centering\arraybackslash}p{#1}}
    \newcolumntype{R}[1]{>{\raggedleft\arraybackslash}p{#1}}
    \usepackage{relsize} 

	\usepackage{hyperref} 

    \usepackage{chicago} 

    \usepackage[]{graphicx}
    \usepackage{amsmath,amssymb,mhchem,booktabs,siunitx,geometry,rotating,threeparttable}
    \usepackage[sort,compress]{cite}
    \usepackage{float}
    \usepackage[caption = false]{subfig}

    \usepackage{mhchem}

	\usepackage{comment}

\usepackage{braket}

  \title{\vspace{-15mm}\fontsize{21pt}{10pt}\selectfont\textbf{The Generalized Occupation-Restricted-Multiple-Active-Space Concept in Multiconfigurational Self-Consistent Field Methods\thanks{Draft manuscript. Please do not cite without the authors' permission.}}}

  \author[1,2]{\large Chen Yang}
  \author[2]{\large Yoshihiro Watanabe}
  \author[2]{\large Haruyuki Nakano}
  \affil[1]{\normalsize Institute of Atomic and Molecular Physics, Sichuan University}
  \affil[2]{\normalsize Department of Chemistry, Graduate School of Science, Kyushu University}
  
\begin{document}
\maketitle



\begin{abstract}
\noindent A novel concept of multiconfigurational self-consistent field method, the generalized occupation-restricted-multiple-active-space (GORMAS) is presented. GORMAS wave functions are defined by substituting the complete active space (CAS) in ORMAS, given a pre- or a post-restriction. The GORMAS approach shows a flexible selection of active space, reduce the ineffective reference configurations from CAS dramatically. Test calculations in molecule or complex systems, \ce{CH2O}, \ce{(H2O)2} molecule, and oxoMn(salen) are presented. They show the GORMAS wave functions achieve the similar accuracy with under 15\% dimension of reference spaces.
\end{abstract}



\doublespacing

\section{Introduction}
Multiconfigurational self-consistent field (MCSCF) methods\cite{Hinze1973, Jorgensen1978, MSGordon1998} have succeed in describing non-dynamical correlation, such as degenerate or quasi-degenerate states, bond breaking\cite{Roos1993}, conical intersections\cite{Andres2003}, transition metal complexes, photochemical mechanism\cite{Liu2015}, actinide chemistry, etc. Moreover, the MCSCF wave functions are usually used as references for subsequent perturbation calculations to include the dynamical correlation, such as the complete active space second-order perturbation theory (CASPT2)\cite{CASPT2-1, CASPT2-2}.

The CAS-SCF method, which is a representative MCSCF method, is most commonly used in the past several decades\cite{casscf}. The most important orbitals (usually near HOMO and LUMO) are selected, and the full configuration interaction (FCI) is constructed using the selected orbitals. By performing FCI in the active space and minimizing the energy for orbital variation, the wave function is determined. The CAS-SCF method has a good description of the electronic structure including static electron correlation. However, the problem is that the number of the determinants in reference space grows in a factorial fashion, and hence the computational cost will dramatically increase when the system becomes larger. The FCI scheme is only possible for some very small systems that contain very few electrons.

The electron configuration is described by the distribution of electrons in the molecular orbitals, that is, the electron occupation number of each orbital. In spite of CAS-SCF method takes full configuration interaction method in the active space, it can provide the best description of electronic structure in the given active space. However, the high computational cost of CAS-SCF is a disadvantage. Therefore, several improvements on building active space, RAS-SCF\cite{RAS}, QCAS-SCF, ORMAS-SCF\cite{ormas} were proposed, to deal with the efficiency problem. Among them, the ORMAS-SCF method is an excellent method, which is a natural extension of CAS-SCF and therefore inherits advantages of CAS-SCF. ORMAS is a construction of variational space that has given specific numbers of occupations, the maximum and minimum number of electrons. In this article, we try to further generalize the concept of ORMAS-SCF method.

The paper is organized as follows: In Sec. \ref{CM}, we briefly introduce the concept and construction schemes of GORMAS. In Sec. \ref{RD}, we calculate some examples to demonstrate the validity and advantages by using GORMAS reference space. In Sec. \ref{Con}, we summarized some conclusions.

\section{Method---Schemes of GORMAS}\label{CM}
\paragraph{Design of GORMAS}
In the present section, we present the GORMAS method, a generalized version of the occupation-restricted-multiple-active-spaces (ORMAS) method. Let us assume the MCSCF wave function $\Psi$ expanded by Slater determinants $\Phi_I$

\begin{equation}
    \Psi =\sum_I C_I \Phi_I
\end{equation}
with

\begin{equation}
    \Phi_I = \prod_i a_{iI}^\dagger \ket{\text{core}}
\end{equation}
where $\ket{\text{core}}$ is the determinant with all the core orbitals filled by electrons.

First, we redefine the ORMAS. In the original paper of Ivanic\cite{Ivanic2003}, the ORMAS is defined as a subspace of the full configuration interaction (FCI) space.   In their definition, the active orbital space, $\varphi_i (i=1,2,\cdots,m_\text{act})$, is first partitioned into several orbital subspaces, $\varphi_{j1} (j=1,2,\cdots,m_1), \varphi_{k2}  (k=1,2,\cdots,m_2),\cdots, \varphi_{lX} (l=1,2,\cdots,m_X)$.  Second, the limitation is placed on the numbers of electrons that can occupy each orbital subspace, specifying the minimum and maximum numbers of electrons ($n_i^\text{min}$ and $n_i^\text{max}$) in orbital subspace i. Then the configuration space is set to include all the possible determinants that satisfy these restrictions. 

\begin{equation}
\text { ORMAS }=\text{CAS} \wedge\left\{n_{i}^{\min } \leq n_{i} \leq n_{i}^{\max }(i=1,2, \cdots, X)\right\}
\end{equation}

This definition is probably the most general and the most natural one as well. Since this definition is based on the restriction of a larger space, i. e. CAS, by imposing some conditions, we refer to this general definition as \emph{post-restriction}.

In addition to the post-restriction, other definitions are also possible for ORMAS. In a previous paper of ours, we presented the quasicomplete active space, which is a product space of several CASs:

\begin{equation}
\text{QCAS}=\bigotimes_{i=1}^{X} \text{CAS}_{i}\left(n_{i}, m_{i}\right)
\end{equation}

Here the CASs in the product are the spaces made from the fixed number of electrons $n_i$ and the set of orbitals in each orbital subspace $i$. If there are no spin constraints, the QCAS is the largest space for the fixed numbers of electrons in each orbital subspace. However, in the actual molecular calculations, the whole space is subject to a fixed spin, and in such a case the QCAS is the sum space of all the possible spin-couplings between each sub-CAS:

\begin{equation}
\text{QCAS}^{\prime}=\bigoplus_{\text{spin}-\text{coupling}} \bigotimes_{i=1}^{X} \text{CAS}_{i}^{S_{i}}\left(n_{i}, m_{i}\right)
\end{equation}
where $S_i$ is the spin of the sub-CAS. Now, if we further take the sum of this QCAS for all combinations of the numbers of electrons in the orbital subspaces that satisfy the restriction of $n_i^\text{min}<n_i<n_i^\text{max}$, we have another definition of ORMAS:

\begin{equation}
\text{ORMAS}=\bigoplus_{C} \bigotimes_{i=1}^{X} \text{CAS}_{i}^{S_{i}}\left(n_{i}, m_{i}\right)
\end{equation}
where $C$ are combinations of electron numbers $(n_1,n_2,\cdots,n_X)$ and spins $(S_1,S_2,\cdots,S_X)$. Since in this definition, the restriction on the electron numbers come first, we refer to this definition as \emph{pre-restriction}.

Then, we generalize the ORMAS concept based on these two definitions to get GORMAS.

Based on the original definition of former by the post-restriction, we can generalize the ORMAS to GORMAS by replacing the CAS before the restriction with a general configuration space (GCS):

\begin{equation}
\text { GORMAS }=\text { GCS } \wedge\left\{n_{i}^{\min } \leq n_{i} \leq n_{i}^{\max }(i=1,2, \cdots, X)\right\}
\end{equation}
where the GCS is a general active space composed of an arbitrary set of Slater determinants.  We refer to this type of GORMAS as GORMAS type 1 (GORMAS-1).

We can also generalize the ORMAS based on the pre-restriction by replacing the sub-CASs in Eq. (6) by the sub-GCSs:

\begin{equation}
\text { GORMAS }=\bigoplus_{C} \bigotimes_{i=1}^{X} \text{GCS}_{i}^{S_{i}}\left(n_{i}, m_{i}\right)
\end{equation}
where the GCSs are the general configuration spaces composed of $n_i$ electrons and $m_i$ active orbitals. We refer to this type of GORMAS as GORMAS type 2 (GORMAS-2).

Now let us illustrate the concept of GORMAS-1 and 2.  We employ the formaldehyde \ce{CH2O} molecule as an example.  Figure \ref{fig:ch2o_orbs} shows some low-lying orbitals of the formaldehyde molecule except for the core orbitals.  If we list the orbital by the occupation sequence, $(3 \text{a}_{1})^{2}(4 \text{a}_{1})^{2}(5 \text{a}_{1})^{2}(6 \text{a}_{1})^{0}(7 \text{a}_{1})^{0}(8 \text{a}_{1})^{0}(9 \text{a}_{1})^{0}(1 \text{b}_{2})^{2}(2 \text{b}_{2})^{2}(3 \text{b}_{2})^{0}(4 \text{b}_{2})^{0}(1 \text{b}_{1})^{2}(2 \text{b}_{1})^{0}(3 \text{b}_{1})^{0}$ is the ground state configuration, where the orbitals are rearranged according to the orbital symmetry. Neglecting the molecular orbital symbols, the ground state configuration determinant is expressed by $22200002200200$ as a shorthand notation. Using these electrons and orbitals, we can create CAS(12e,14o), which means a complete active space with 12 active electrons and 14 active orbitals, as in the usual convention. Dividing the active orbitals into 7, 4, and 3 orbitals according to the orbital symmetry and limiting the number of electrons in each orbital group to 6, 3-4, and 2-3 orbitals, respectively, we have ORMAS$\left(7 \frac{6}{6}, 4 \frac{3}{4}, 3 \frac{2}{3}\right)$. Here, ORMAS$\left(m_{1} \frac{n_{1}^{\min }}{n_{1}^{\max }}, m_{2} \frac{n_{2}^{\min }}{n_{2}^{\max }}, m_{3} \frac{n_{3}^{\min }}{n_{3}^{\max }}\right)$ represents ORMAS constructed by several sub-CAS$_{n}$ subspaces that defined with minimum $\left(n_{n}^{\min }\right)$ and maximum $\left(n_{n}^{\max }\right)$ electron occupations in $m_{n}$ orbitals.

Replacing the CAS(12e,14o) with another space yields a GORMAS in post-restriction type (GORMAS-1). For a simple example, if we use the ground state configuration plus singles and doubles space for a simple example, we have

GORMAS-1$\left(7 \frac{6}{6}, 4 \frac{3}{4}, 3 \frac{2}{3}\right) \mid 22200002200200$+SD
where GORMAS-1$\left(m_{1} \frac{n_{1}^{\min }}{n_{1}^{\max }}, m_{2} \frac{n_{2}^{\min }}{n_{2}^{\max }}, m_{3} \frac{n_{3}^{\min }}{n_{3}^{\max }}\right)$ $\mathrm{parent\_determinant}+n$ ex represents the same ORMAS above intersection with a maximum $n$ electrons excited space generated from a $\mathrm{parent\_determinant}$.

Then, replacing sub-CAS in ORMAS$\left(7 \frac{6}{6}, 4 \frac{3}{4}, 3 \frac{2}{3}\right)$ with other spaces yields a GORMAS in pre-restriction type. If we use $2220000$+SDTQ in the $a_{1}$ orbital space, $2200$+SD in the b$_{2}$ orbital space, and $200+$SD in the b$_{1}$ orbital space instead of the sub-CAS$_{n}$, we have
GORMAS-2(4/2/2) $2220000,2200,200$, where GORMAS-2$\left(n_{1} / n_{2} / \cdots\right)$ pdet$_{1}$, pdet$_{2}, \cdots$ represents a direct sum space constructed from the general configuration spaces generated with maximum $n$ electrons excitations from pdet$_{n}$.

\begin{figure}[]
\centering
\includegraphics[width=0.8\linewidth]{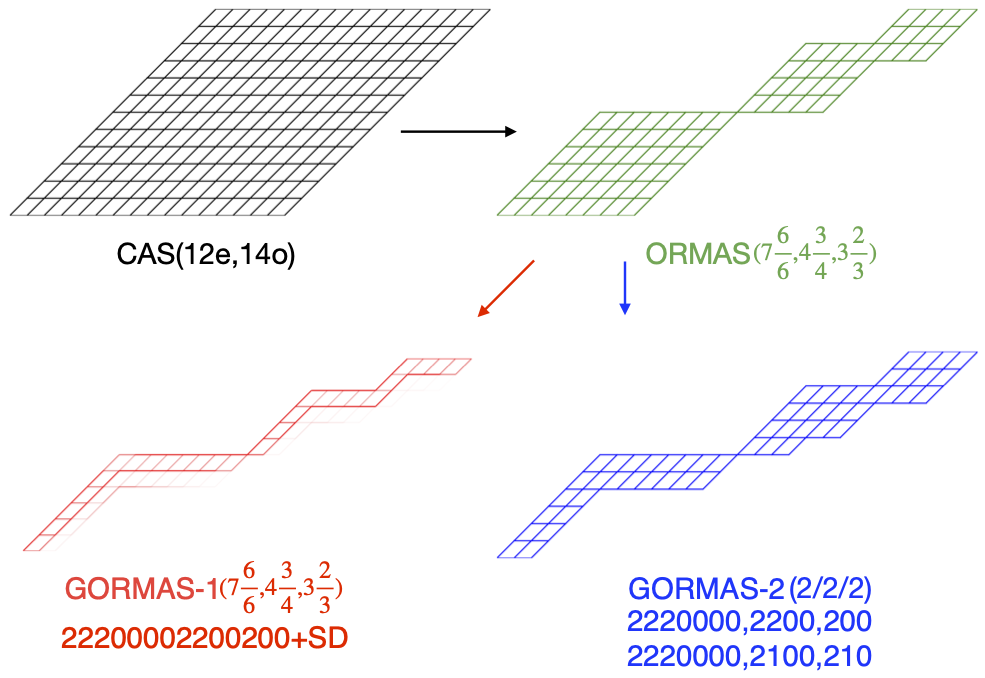}
\caption{Graph representation of \ce{CH2O} electronic active space.}\label{fig:VSCnew}
\end{figure}

Figure \ref{fig:VSCnew} shows the conceptual graph representations of the example active spaces CAS(12e,14o), ORMAS$\left(7 \frac{6}{6}, 4 \frac{3}{4}, 3 \frac{2}{3}\right)$, GORMAS-1$\left(7 \frac{6}{6}, 4 \frac{3}{4}, 3 \frac{2}{3}\right) \mid 22200002200200$+SD, and GORMAS-2(2/2/2) $2220000,2200,200$. The graph representation is based on GAMESS package and Duch's book.\cite{Schmidt1993,duch1986grms}

\section{Results and Discussion}\label{RD}
To demonstrate the performance of the GORMAS-SCF method, we calculated formaldehyde, water dimer, and the oxoMn(salen) complex and compared the results with those of other MCSCF methods. All the calculations were performed with a modified version of the GAMESS program package.\cite{Schmidt1993} 

\subsection{Ground and first excited states of formaldehyde, \ce{CH2O}}
Formaldehyde has an excitation from the non-bonding lone-pair orbital to the anti-bonding $\pi$ orbital, which corresponds to the lowest singlet excited state. We calculated the ground and first excited states using the GORMAS-SCF method with several GORMAS designs and compared the results with those of the ORMAS-SCF method. The molecular structure was taken from experimental data\cite{YAMADA1971}. The basis set used was the correlation-consistent polarized valence triple-zeta (cc-pVTZ) set.\cite{Dunning1989,Balabanov2005} The results are summarized in Table \ref{tab:CH2O}.

\begin{figure}
\centering
\subfloat[orb 3: $3a_1$]{\includegraphics[width = 0.15\linewidth]{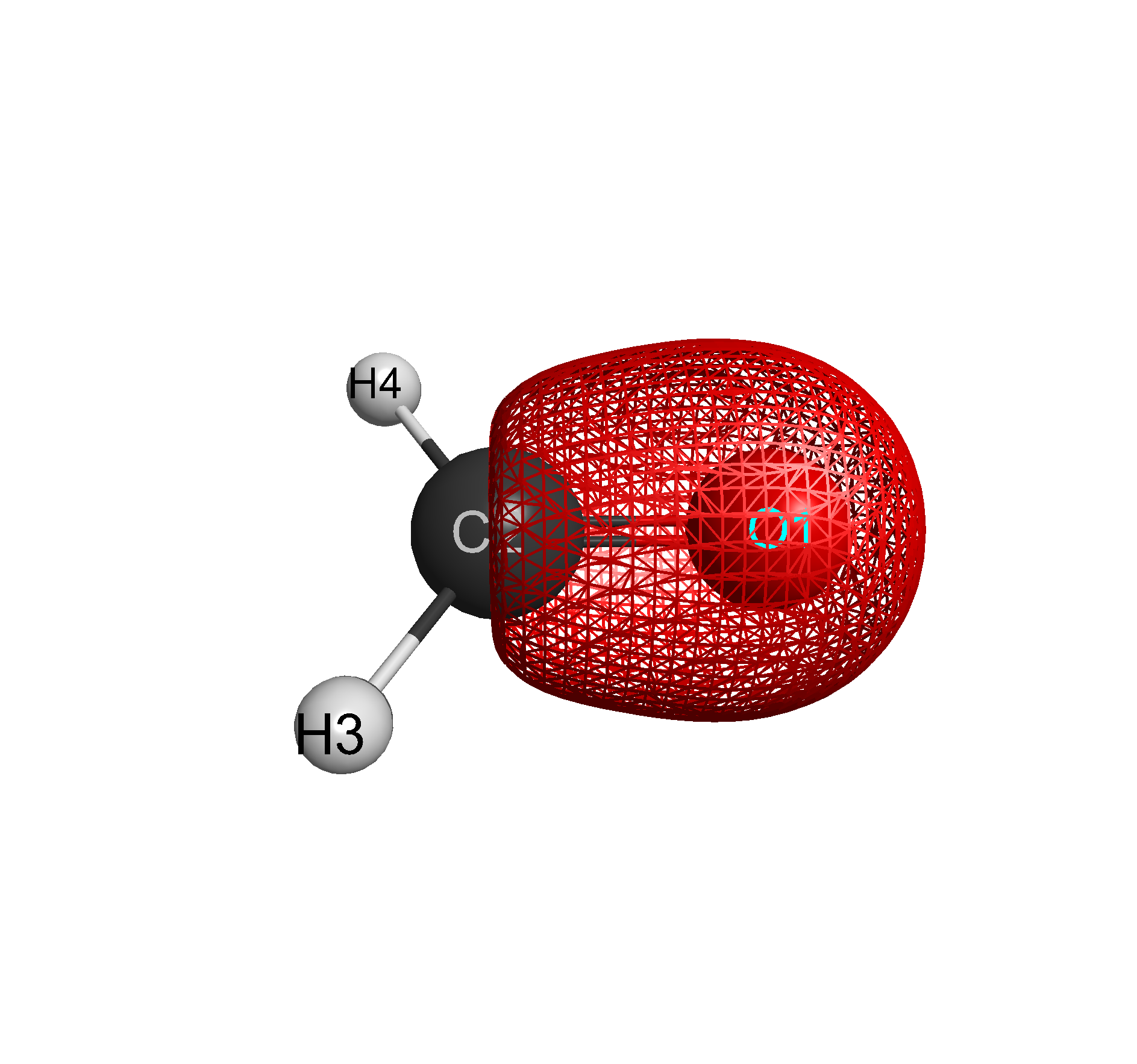}}\hfil
\subfloat[orb 4: $4a_1$]{\includegraphics[width = 0.15\linewidth]{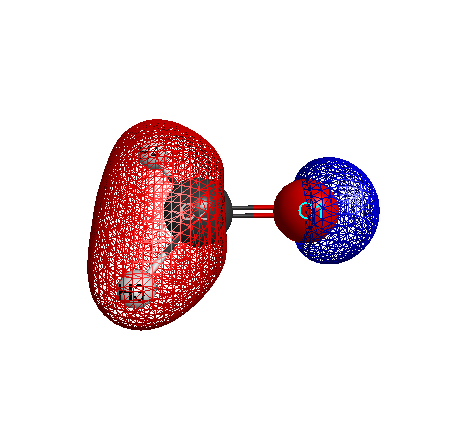}}\hfil
\subfloat[orb 5: $5a_1$]{\includegraphics[width = 0.15\linewidth]{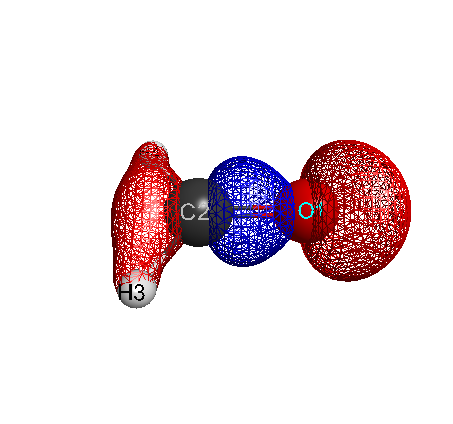}}\hfil
\subfloat[orb 6: $6a_1$]{\includegraphics[width = 0.15\linewidth]{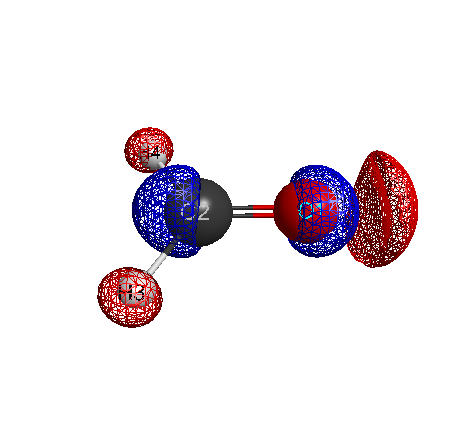}}\\
\subfloat[orb 7: $7a_1$]{\includegraphics[width = 0.15\linewidth]{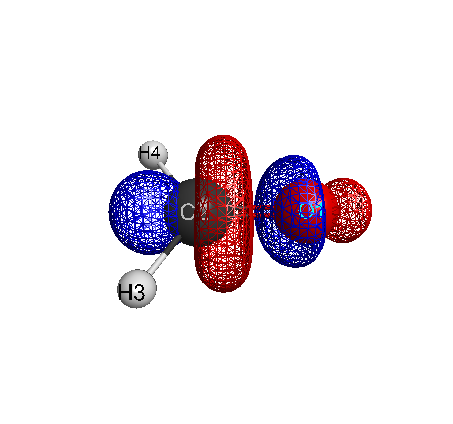}}\hfil
\subfloat[orb 8: $8a_1$]{\includegraphics[width = 0.15\linewidth]{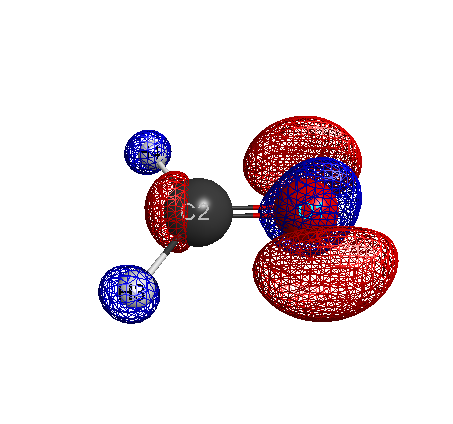}}\hfil
\subfloat[orb 9: $9a_1$]{\includegraphics[width = 0.15\linewidth]{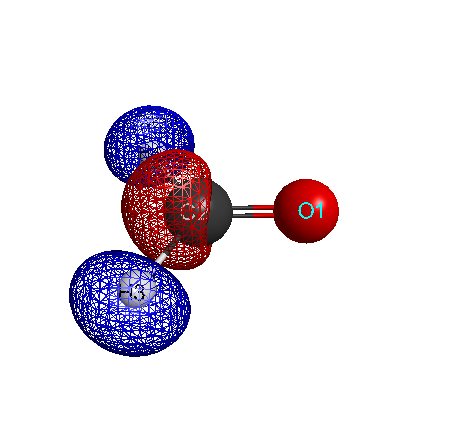}}\\
\subfloat[orb 10: $1b_2$]{\includegraphics[width = 0.15\linewidth]{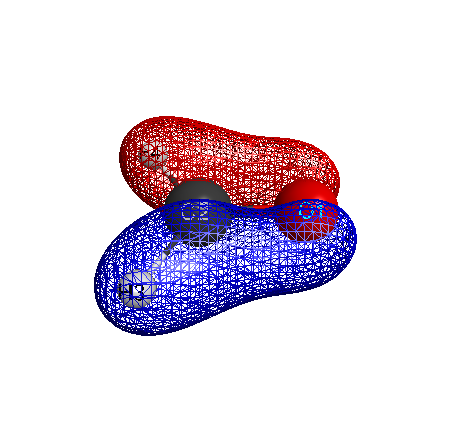}}\hfil
\subfloat[orb 11: $2b_2$]{\includegraphics[width = 0.15\linewidth]{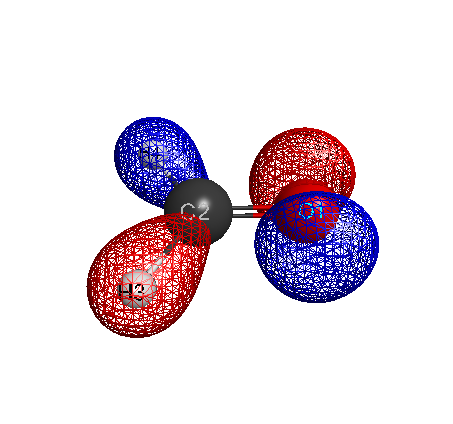}}\hfil
\subfloat[orb 12: $3b_2$]{\includegraphics[width = 0.15\linewidth]{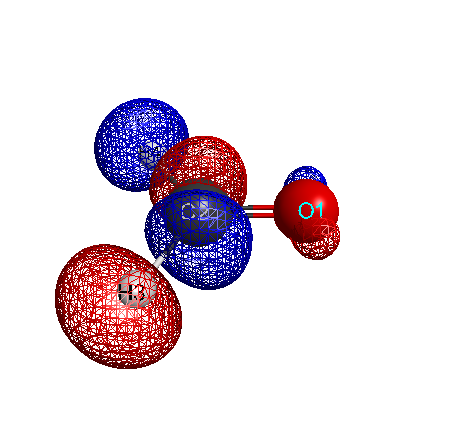}}\hfil
\subfloat[orb 13: $4b_2$]{\includegraphics[width = 0.15\linewidth]{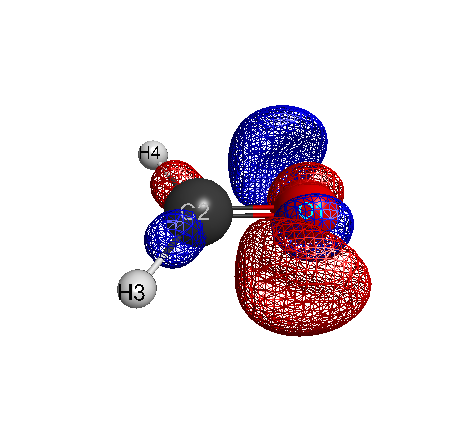}}\\
\subfloat[orb 14: $1b_1$]{\includegraphics[width = 0.15\linewidth]{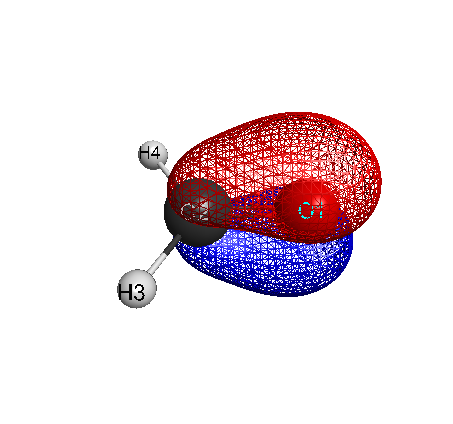}}\hfil
\subfloat[orb 15: $2b_1$]{\includegraphics[width = 0.15\linewidth]{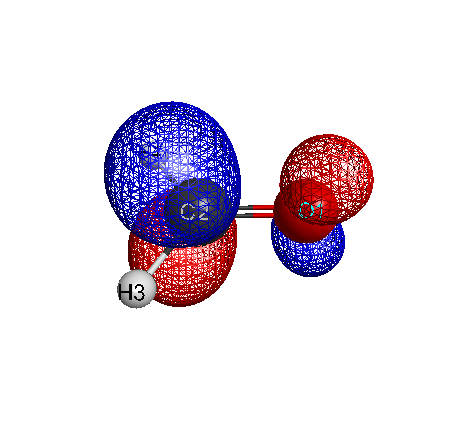}}\hfil
\subfloat[orb 16: $3b_1$]{\includegraphics[width = 0.15\linewidth]{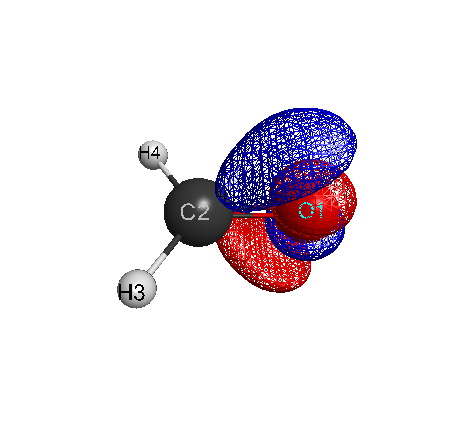}}
\caption[Orbitals of \ce{CH2O} active space.]{Orbitals of \ce{CH2O} active space. (i) orb 11 is the HOMO, and (m) orb 15 is the LUMO.}
\label{fig:ch2o_orbs}
\end{figure}

The active orbitals used are 3a$_1$-9a$_1$, 1b$_2$-4b$_2$, and 1b$_1$–3b$_1$, shown as examples in the previous section. The first group 3a$_1$-9a$_1$ correspond to $\sigma$ bonding (orbs. 3, 4, and 5 in Fig. \ref{fig:ch2o_orbs}) and $\sigma$ anti-bonding (orbs. 6, 7, 8, and 9) orbitals. the second group 1b$_2$-4b$_2$ in Fig. \ref{fig:ch2o_orbs} non-bonding (orbs. 11 and 13), and $\sigma$ anti-bonding orbitals in Fig. \ref{fig:ch2o_orbs} (orb. 12), and the third group 1b$_1$-3b$_1$ to $\pi$ bonding (orb. 14) and anti-bonding (orbs. 15 and 16) orbitals. If we use all these orbitals and 12 electrons, the dimension of the full space, CAS(12e,14o) is 9 018 009. For this molecule, we used the ORMAS that was constructed using three groups of orbitals with same symmetry ($\text{ORMAS}\left(7 \frac{6}{6}, 4 \frac{3}{4}, 3 \frac{2}{3}\right)$), as the reference space for the calculations. It has a configuration dimension of 1 868 566. For GORMAS-1 calculations, we tried two spaces with two and four excitations from a parent determinant $2220000 2200 200$, namely GORMAS-1$\left(7 \frac{6}{6}, 4 \frac{3}{4}, 3 \frac{2}{3}\right)$ $2220000 2200 200$/2ex and /4ex. For the GORMAS-1/2ex with a dimension 1 075, the differences of the ground and excited state energies from the ORMAS values were 0.0316 and 0.0786 hartree, respectively. These energies due to that small space have a rather large error, and the errors were not even, i. e., the ground and excited states were not balanced. The inclusion of triple and quadruple excitations improved the energies and balance. For the GORMAS-1/4ex with a dimension 79 104, the differences of the ground and excited state energies from the ORMAS values were 0.0001 and 0.0026 hartree, respectively. As a result, the difference from the ORMAS values in excitation energy was reduced from 0.8588 eV to 0.069eV.

The same partitioning of active orbitals as ORMAS and GORMAS-1 was used for GORMAS-2. The GORMAS-2 was constructed as the sum space of two product spaces $2220000$+SD $\otimes$ $2200$+SD $\otimes$ $200$+SD and $2220000$+SD $\otimes$ $2100$+SD $\otimes$ $210$+SD. The dimension of this GORMAS-2 was 251 889; GORMAS-2 reduced about 87\% size of dimension. The differences of the ground and excited state energies from the ORMAS values were 0.0018 and 0.0014 hartree, respectively, and the difference in excitation energy was $-$0.012  eV.

\begin{table}
\centering
\caption[Energies of the ground (S$_0$) and excited (S$_1$) states for \ce{CH2O}.]{Energies of the ground (S$_0$) and excited (S$_1$) states for \ce{CH2O}. (in hartree)}
\label{tab:CH2O}
\begin{tabular}{lrcccc}
\hline
\multicolumn{1}{c}{Method}                                                            & \multicolumn{1}{c}{Dimension}  & 
S$_0$           & S$_1$           & Excitation Energy (eV) \\ \hline
ORMAS$(7\frac{6}{6},4\frac{3}{4},3\frac{2}{3})$                                       & 1868566                        & --114.104976 & --113.950035 & 4.216                  \\
GORMAS-1$^a(7\frac{6}{6},4\frac{3}{4},3\frac{2}{3})$                                   & 1075                           & --114.073415 & --113.871419 & 5.497                  \\
22200002200200/2ex                                                                    &                                &              &              &                        \\
GORMAS-1$^a$ + PT2                                                                       & 1075                           & --114.3137   & --114.1742   & 3.795                 \\
GORMAS-1$^b(7\frac{6}{6},4\frac{3}{4},3\frac{2}{3})$                                   & 79104                          & --114.104869 & --113.947407 & 4.285                  \\
22200002200200/4ex                                                                    &                                &              &              &                        \\
GORMAS-1$^b$ + PT2                                                                       & 79104                          & --114.3146   & --114.1691   & 3.961                 \\
GORMAS-2(2/2/2)                                                                        & 251889                         & --114.103174 & --113.948665 & 4.204                  \\
2220000,2200,200                                                                      &                                &              &              &                        \\
2220000,2100,210                                                                      &                                &              &              &                        \\ \hline
\end{tabular}
\end{table}

\subsection{Water dimer \ce{(H2O)2}}
\begin{figure}[]
\centering
\includegraphics[width=0.5\linewidth]{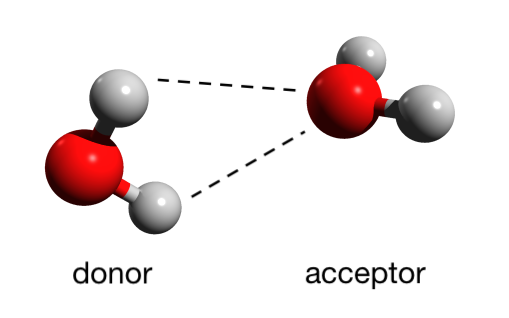}
\caption{Model of bifurcated \ce{(H2O)2}.}\label{fig:water_dimer}
\end{figure}

The second example was the water dimer, which has several stable structures. According to the literature, there are three types of dimer geometries\cite{Lane2013,Shank2009}. As an example of GORMAS-SCF calculation, we optimized the bifurcated water dimer which includes a donor water and an acceptor water molecules shown in Figure \ref{fig:water_dimer}. The basis set used was the cc-pVTZ set\cite{Dunning1989,Balabanov2005}. The active orbitals were the symmetric and antisymmetric $\sigma$ bonding orbitals, the lone-pair orbital, the two symmetric and two antisymmetric $\sigma$ antibonding orbitals of each water molecule, for a total 14 orbitals. The 12 valence electrons, excluding the oxygen 2s electrons, were used to make CAS, ORMAS and GORMAS. The results are summarized in Table \ref{tab:H2O}.

The reference for comparison was CAS(12e,14o), where 12 valence electrons were distributed to the 14 active orbitals above. Table \ref{tab:H2O} shows that the water dimer has quasidegenerate excited states, the first excited state S$_1$ and the second excited state S$_2$ are close to each other with excitation energies of 8.278 eV and 8.468 eV, respectively. The S$_1$ has the excitation mainly excited from the non-bonding orbital of the acceptor water to the $\sigma$ anti-bonding orbital of the acceptor water, while the S$_2$ has mixed excitation modes including an excitation from the non-bonding orbital of the donor water to the $\sigma$ anti-bonding orbital of the donor water and an excitation from the non-bonding orbital of the donor water to the $\sigma$ anti-bonding orbital of the acceptor water.

\begin{figure}
\centering
\subfloat[orb 5: $1\sigma_\text{A}$]{\includegraphics[width = 0.15\linewidth]{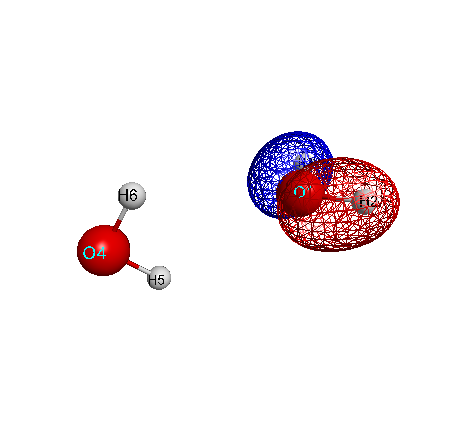}}\hfil
\subfloat[orb 6: $2\sigma_\text{A}$]{\includegraphics[width = 0.15\linewidth]{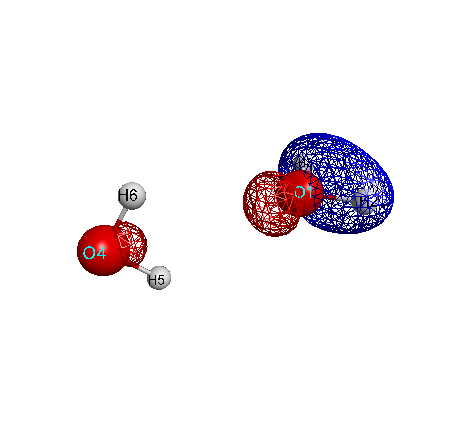}}\hfil
\subfloat[orb 7: n$_\text{A}$]{\includegraphics[width = 0.15\linewidth]{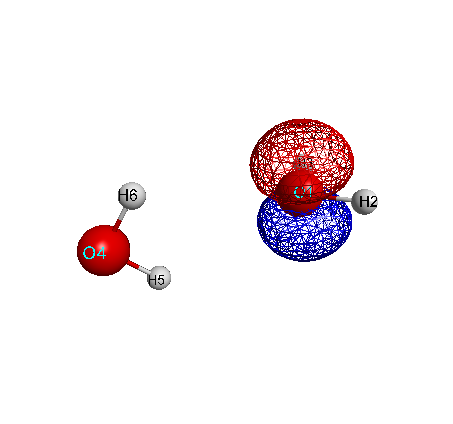}}\hfil
\subfloat[orb 8: $1\sigma_\text{A}^*$]{\includegraphics[width = 0.15\linewidth]{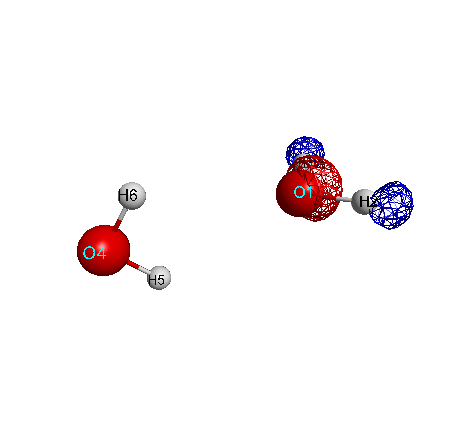}}\\
\subfloat[orb 9: $2\sigma_\text{A}^*$]{\includegraphics[width = 0.15\linewidth]{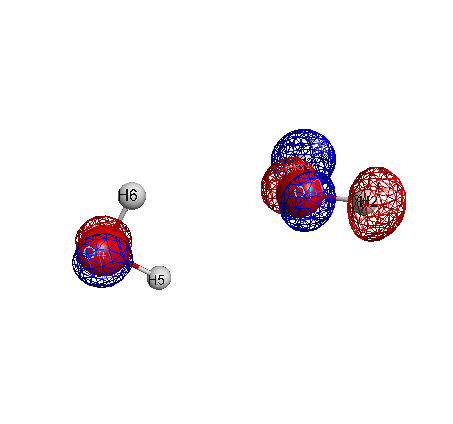}}\hfil
\subfloat[orb 10: $3\sigma_\text{A}^*$]{\includegraphics[width = 0.15\linewidth]{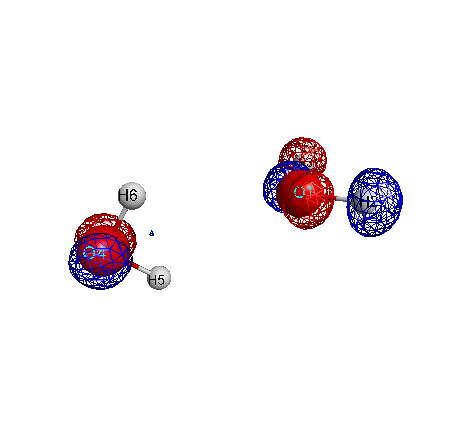}}\hfil
\subfloat[orb 11: $4\sigma_\text{A}^*$]{\includegraphics[width = 0.15\linewidth]{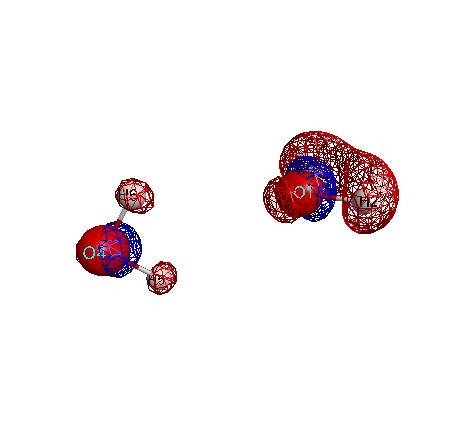}}\\
\subfloat[orb 12: $1\sigma_\text{D}$]{\includegraphics[width = 0.15\linewidth]{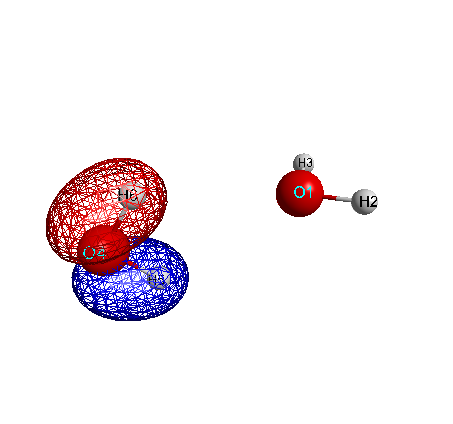}}\hfil
\subfloat[orb 13: $2\sigma_\text{D}$]{\includegraphics[width = 0.15\linewidth]{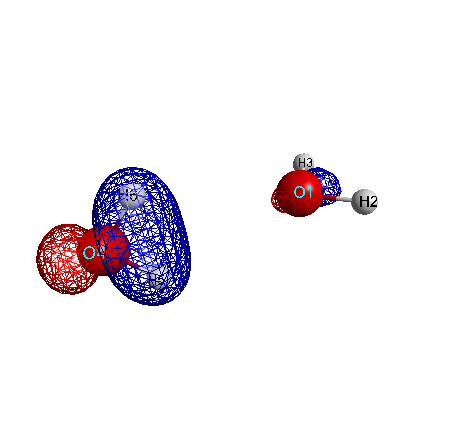}}\hfil
\subfloat[orb 14: n$_\text{D}$]{\includegraphics[width = 0.15\linewidth]{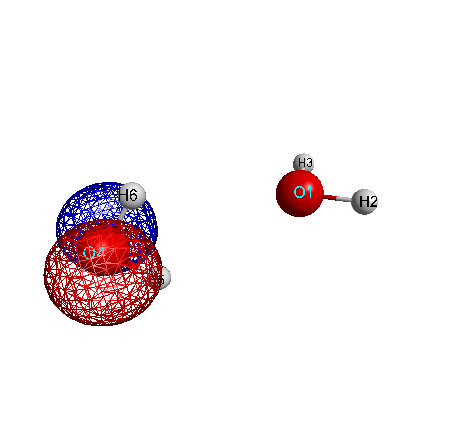}}\hfil
\subfloat[orb 15: $1\sigma_\text{D}^*$]{\includegraphics[width = 0.15\linewidth]{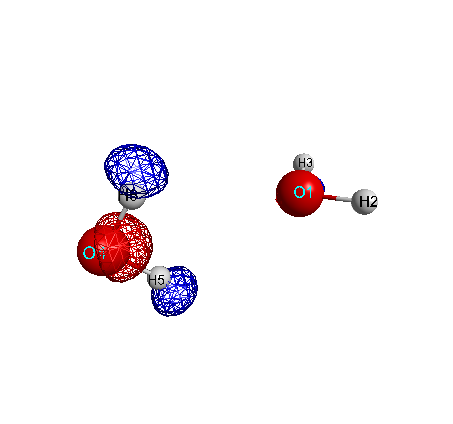}}\\
\subfloat[orb 16: $2\sigma_\text{D}^*$]{\includegraphics[width = 0.15\linewidth]{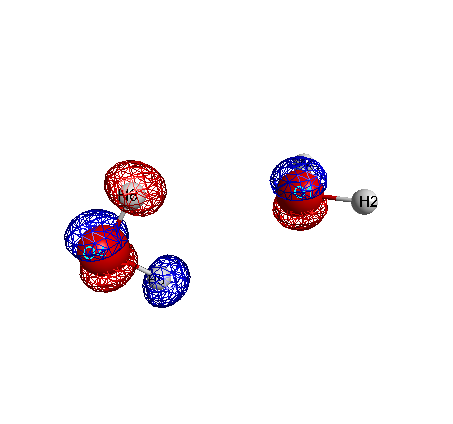}}\hfil
\subfloat[orb 17: $3\sigma_\text{D}^*$]{\includegraphics[width = 0.15\linewidth]{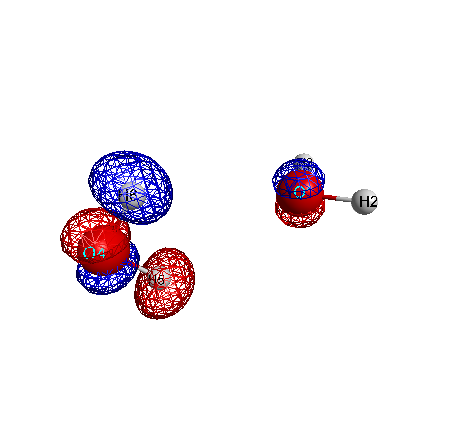}}\hfil
\subfloat[orb 18: $4\sigma_\text{D}^*$]{\includegraphics[width = 0.15\linewidth]{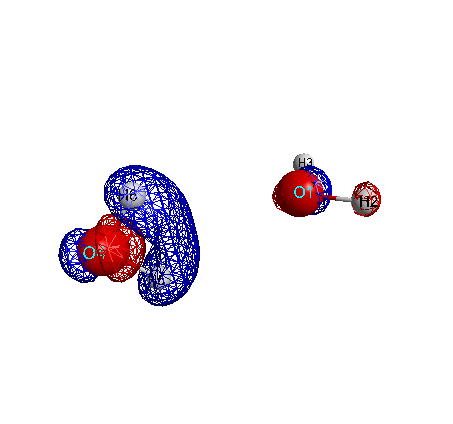}}
\caption[Orbitals of \ce{(H2O)2} active space.]{Orbitals of \ce{(H2O)2} active space. (j) orb 14 is the HOMO, and (d) orb 8 is the LUMO.}
\label{fig:h2o_orbs}
\end{figure}

GORMAS-1$(7\frac{4}{8},7\frac{4}{8})$ was constructed from two orbital groups simply partitioned by the donor water or acceptor water orbitals. The parent determinants were two one electron excited determinants: one is an internal excitation in the acceptor water orbital group and the other is excitation between two orbital groups. This GORMAS-1 reduced about 99.8\% size of dimension, with the largest error 0.0134 hartree (0.3644 eV) in S$_2$ energy, and 0.046 eV energy difference in excitation energy.

GORMAS-2 calculation contains three subspaces, the first subspace includes $\sigma$ bonding and virtual orbitals (Figure \ref{fig:h2o_orbs}, orbs 5, 6, 9, 10) of the donor water, the second subspace includes $\sigma$ bonding and virtual orbitals of the acceptor water (Figure \ref{fig:h2o_orbs}, orbs 12, 13, 16, 17), the third subspace includes all non-bonding orbitals of donor and acceptor waters (Figure \ref{fig:h2o_orbs}, orbs 7, 8, 11, 14, 15, 18). The results show that the GORMAS-2 calculation reduced about 95.6\% size of dimension, with the largest error 0.0243 hartree (0.6621 eV) in S$_2$ energy, and 0.011 eV in excitation energy.

In this case, both GORMAS spaces have reduced the dimension dramatically. The largest energy difference to CAS is 0.66 eV. The most important excitation happened in localized water HOMO--1 (\ce{n_A}), and one excitation from donor water lone pair HOMO (\ce{n_D}) to the acceptor water LUMO (\ce{1\sigma^*_A}). The optimized orbitals in active space are showed in Figure \ref{fig:h2o_orbs}.

\begin{sidewaystable}
\centering
\caption[Energies of ground (S$_0$) and excited (S$_1$ and S$_2$) states for \ce{(H2O)2}.]{Energies of GS, S$_1$ and S$_2$ for \ce{(H2O)2}. (in hartree)}\label{tab:H2O}
\begin{tabular}{lrcccc}
\toprule
\multicolumn{1}{c}{Method} & \multicolumn{1}{c}{Dimension} & MCSCF       & Excitation Energy (eV) & Excitation Mode & Percentage(\%) \\
\midrule
CAS(12e,14o)               & 9018009                  & --152.297799 &        &          &                \\
                           &                          & --151.993581 & 8.278  &   1\ce{^1A$^\prime$$^\prime$(n_A -> 1\sigma^*_A)}       &   87.4             \\
                           &                          & --151.986617 & 8.468  &   2\ce{^1A$^\prime$(n_D -> 1\sigma^*_D)}       &   52.9             \\
                           &                          &             &        &   2\ce{^1A$^\prime$(n_D -> 1\sigma^*_A)}       &   33.9             \\
CAS(12e,14o) + PT2         & 9018009                  & --152.637685 &        &          &                \\
                           &                          & --152.334940 & 8.238  &   1\ce{^1A$^\prime$$^\prime$(n_A -> 1\sigma^*_A)}       &                \\
                           &                          & --152.331657 & 8.327  &   2\ce{^1A$^\prime$(n_D -> 1\sigma^*_D)}       &                \\
                           &                          &             &        &   2\ce{^1A$^\prime$(n_D -> 1\sigma^*_A)}       &                \\
GORMAS-1$(7\frac{4}{8},7\frac{4}{8})$                    & 14929                    & --152.637497 &        &          &                \\
22200002220000             &                          & --152.333692 & 8.267  &   1\ce{^1A$^\prime$$^\prime$(n_A -> 1\sigma^*_A)}       &   93.5             \\
22110002220000             &                          & --152.329744 & 8.374  &   2\ce{^1A$^\prime$(n_D -> 1\sigma^*_D)}       &    16.3            \\
22210002210000             &                          &             &        &   2\ce{^1A$^\prime$(n_D -> 1\sigma^*_A)}       &    77.1            \\
22200002211000/2ex         &                          &             &        &          &                \\
GORMAS-1$(7\frac{4}{8},7\frac{4}{8})$ + PT2              & 14929                    & --152.286097 &        &          &                \\
22200002220000             &                          & --151.980358 & 8.320  &   1\ce{^1A$^\prime$$^\prime$(n_A -> 1\sigma^*_A)}       &                \\
22110002220000             &                          & --151.973226 & 8.514  &   2\ce{^1A$^\prime$(n_D -> 1\sigma^*_D)}       &               \\
22210002210000             &                          &             &        &   2\ce{^1A$^\prime$(n_D -> 1\sigma^*_A)}       &                \\
22200002211000/2ex         &                          &             &        &          &                \\
GORMAS-2(2/2/4)             & 396495                   & --152.275555 &        &          &                \\
2200,2200,220000           &                          & --151.969733 & 8.322  &   1\ce{^1A$^\prime$$^\prime$(n_A -> 1\sigma^*_A)}       &   90.6             \\
                           &                          & --151.962286 & 8.524  &   2\ce{^1A$^\prime$(n_D -> 1\sigma^*_D)}       &    57.8           \\
                           &                          &             &        &   2\ce{^1A$^\prime$(n_D -> 1\sigma^*_A)}       &    33.2           \\
\bottomrule
\end{tabular}
\end{sidewaystable}

\subsection{The oxoMn(salen) complex}
\begin{figure}[]
\centering
\includegraphics[width=0.5\linewidth]{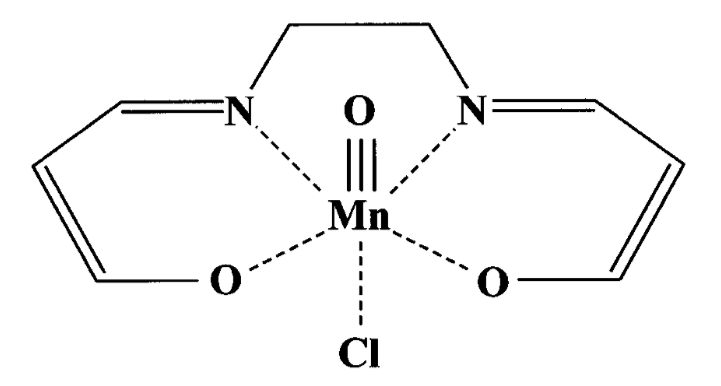}
\caption{Model of oxoMn(salen).}\label{fig:salen}
\end{figure}

The original ORMAS literature used a complex molecule oxoMn(salen) as an application example of the computational method.\cite{Ivanic2003} We also adopted this system as an example. The basis used was the Gaussian-type 6-31g(d)\cite{631gd1,631gd2,631gd3,631gd4,631gd5}, and the same geometry in the reference literature \cite{Ivanic2004} was used.  A reference space of this system is CAS(12e,11o) where five doubly occupied bonding orbitals, $\sigma(\text{O}_\text{ax})$, $\pi_1(\text{O}_\text{ax})$, $\pi_2(\text{O}_\text{ax})$, $\pi_1$, $\pi_2$, one non-bonding orbital, $3d_{x^2-y^2}$(Mn), and five anti-bonding orbitals, $\sigma^*(\text{O}_\text{ax})$, $\pi_1^*(\text{O}_\text{ax})$, $\pi_2^*(\text{O}_\text{ax})$, $\pi_1^*$, $\pi_2^*$ are considered. Symbol O$_\text{ax}$ indicates the O atom in the axial position with respect to the Mn atom.\cite{Ma2011}

\begin{figure}
\centering
\subfloat[orb 64: $\pi_1(\text{O}_\text{ax})$]{\includegraphics[width = 0.21\linewidth]{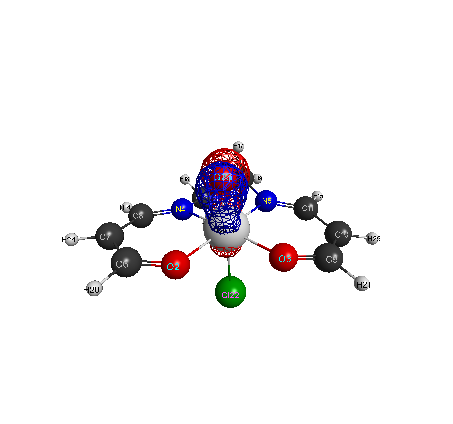}}\hfil
\subfloat[orb 65: $\pi_1$]{\includegraphics[width = 0.21\linewidth]{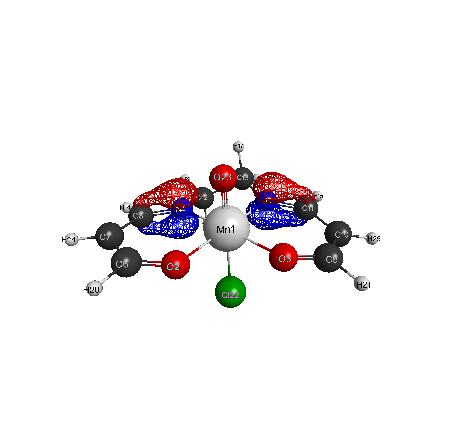}}\hfil
\subfloat[orb 66: $\pi_1^*(\text{O}_\text{ax})$]{\includegraphics[width = 0.21\linewidth]{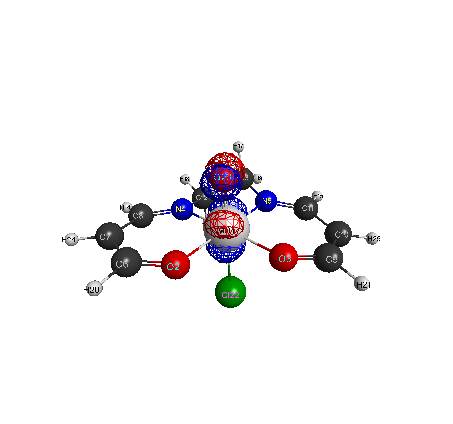}}\hfil
\subfloat[orb 67: $\pi_1^*$]{\includegraphics[width = 0.21\linewidth]{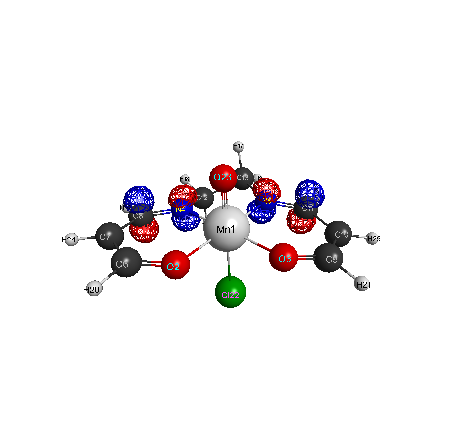}}\\
\subfloat[orb 68: $\pi_2(\text{O}_\text{ax})$]{\includegraphics[width = 0.21\linewidth]{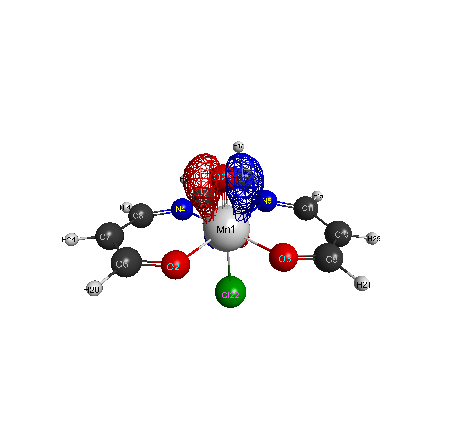}}\hfil
\subfloat[orb 69: $\pi_2$]{\includegraphics[width = 0.21\linewidth]{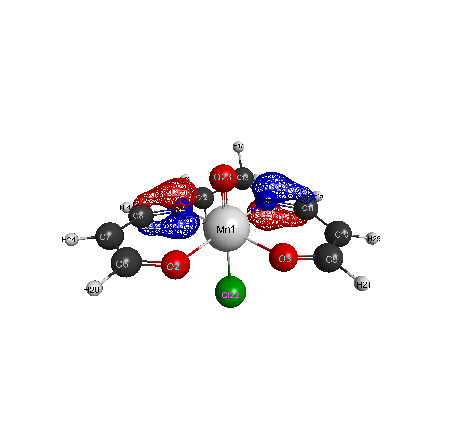}}\hfil
\subfloat[orb 70: $\pi_2^*(\text{O}_\text{ax})$]{\includegraphics[width = 0.21\linewidth]{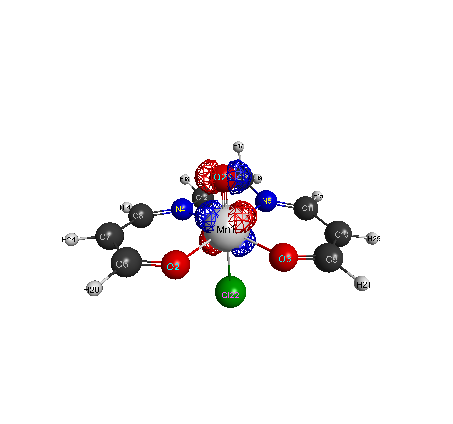}}\hfil
\subfloat[orb 71: $\pi_2^*$]{\includegraphics[width = 0.21\linewidth]{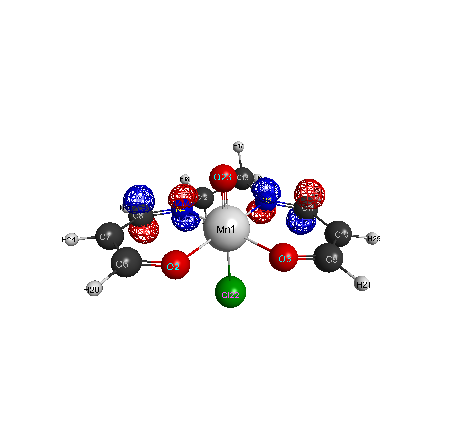}}\\
\subfloat[orb 72: $\sigma(\text{O}_\text{ax})$]{\includegraphics[width = 0.21\linewidth]{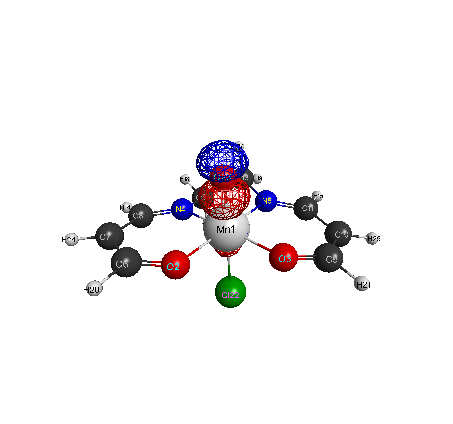}}\hfil
\subfloat[orb 73: $\sigma^*(\text{O}_\text{ax})$]{\includegraphics[width = 0.21\linewidth]{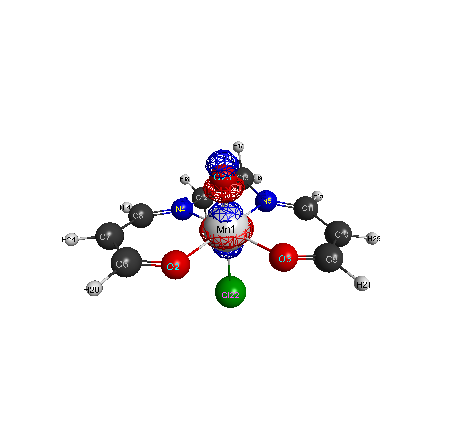}}\hfil
\subfloat[orb 74: $3d_{x^2-y^2}$(Mn)]{\includegraphics[width = 0.21\linewidth]{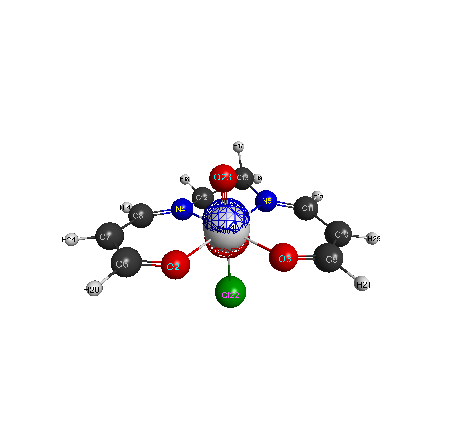}}
\caption{Orbitals of oxoMn(salen) active space.}
\label{fig:Mn_orbs}
\end{figure}

ORMAS$(4,4,3)$ divide the active orbitals of CAS(12e,11o) into three orbital subspaces, where the first subspace consists of $\pi_1(\text{O}_\text{ax})$, $\pi_1$, $\pi_1^*(\text{O}_\text{ax})$, $\pi_1^*$, the second subspace consists of $\pi_2(\text{O}_\text{ax})$, $\pi_2$, $\pi_2^*(\text{O}_\text{ax})$, $\pi_2^*$, the third subspace consists of $\sigma(\text{O}_\text{ax})$, $\sigma^*(\text{O}_\text{ax})$, $3d_{x^2-y^2}$(Mn).(Figure \ref{fig:Mn_orbs}) From the results in Table \ref{tab:salen}, ORMAS$(4\frac{4}{4},4\frac{4}{4},3\frac{4}{4})$, where no electron excitation is allowed among subspaces case, has an error about 0.50 eV when compared with the CAS. In this ORMAS, the number of reference determinants is reduced from 213444 to 23394. However, ORMAS$(4\frac{2}{6},4\frac{2}{6},3\frac{2}{6})$, which has up to two electrons inter-space excitations allowed, shows almost the same results to CAS.

Similarly, GORMAS-1$(4\frac{2}{6},4\frac{2}{6},3\frac{2}{6})$ $22002200220/4$ex is a subset of OR-\linebreak MAS$(4\frac{4}{4},4\frac{4}{4},3\frac{4}{4})$. The results look like a compromised plan of two ORMAS calculations above. Furthermore, GORMAS-2(2/2/2) is an subset of ORMAS$(4\frac{4}{4},4\frac{4}{4},3\frac{4}{4})$, resulted a smaller dimension with a larger error 0.6133 eV. If the inter-space excitation, that is, electrons excited between two subspaces were in consideration, GORMAS-2(2/2/2) in the last row of Table \ref{tab:salen} shows the results are unremarkably same to CAS.

For a complex system, the GORMAS reproduced the similar results to ORMAS. Although the absolute energy difference to CAS seems large, the relative error is still fairly small (0.03\%). However, current construction of subspaces does not have the benefits both in dimension and accuracy.

\begin{table}
\centering
\caption[Ground state energies of oxoMn(salen).]{Ground state energies of oxoMn(salen). (in hartree)}\label{tab:salen}
\begin{tabular}{lrcc}
\toprule
\multicolumn{1}{c}{Method} & \multicolumn{1}{c}{Dimension} & Total Energy         & $\Delta E$ (eV)              \\
\midrule
CAS(12e,11o)               & 213444                   & --2251.431090         & 0.0000               \\
ORMAS$(4\frac{2}{6},4\frac{2}{6},3\frac{2}{6})$                      & 192378                   & --2251.431083         & 0.0002               \\
ORMAS$(4\frac{4}{4},4\frac{4}{4},3\frac{4}{4})$                      & 23394                    & --2251.412660         & 0.5015               \\
GORMAS-1$(4\frac{2}{6},4\frac{2}{6},3\frac{2}{6})$                    & 42485                    & --2251.425038         & 0.1647               \\
22002200220/4ex            &                          &                      &                      \\
GORMAS-2(2/2/2)             & 13203                    & --2251.408552         & 0.6133               \\
2200,2200,220              &                          &                      &                      \\
GORMAS-2(2/2/2)             & 187511                   & --2251.430758         & 0.0090               \\
2200,2200,220              &                          & \multicolumn{1}{l}{} & \multicolumn{1}{l}{} \\
2100,2100,210              &                          & \multicolumn{1}{l}{} & \multicolumn{1}{l}{} \\
2210,2210,221              &                          & \multicolumn{1}{l}{} & \multicolumn{1}{l}{} \\
2000,2000,200              & \multicolumn{1}{l}{}     & \multicolumn{1}{l}{} & \multicolumn{1}{l}{} \\
2220,2220,222              & \multicolumn{1}{l}{}     & \multicolumn{1}{l}{} & \multicolumn{1}{l}{} \\
\bottomrule
\end{tabular}
\end{table}

\section{Conclusions}\label{Con}
In this paper, a novel construction of active space in the MCSCF method, namely, the generalized ORMAS method, was proposed.

In the calculation of \ce{CH2O}, GORMAS reduced at least 87\% size of reference space to the CAS-SCF, gave the same excitation picture and energetic information 
under error in 0.1 eV. In the calculation of \ce{(H2O)2}, GORMAS reduced more than 95\% dimension to the CAS-SCF, and with largest 0.011 eV error in excitation energies. In the case of oxoMn(salen), GORMAS-1 has reduced over 80\% determinants space to the CAS with an error about 0.5 eV, which is a relative 0.02\% error to the CAS-SCF result.

Several examples of practical applications showed that the selection of parent determinants is important for the GORMAS space construction. However, they also showed that the ground state plus several single excitations (which could be obtained by CIS calculation) with four-electron-excitation level gives a generally good computational results. Overall, our results give a consistent accuracy to CAS or ORMAS calculation with smaller dimension.

\newpage
\singlespacing


    \bibliographystyle{references/jcp}
	\bibliography{references/biblio} 

\end{document}